# New Virial Equation of State for Hard-Disk Fluids

Jianxiang Tian [1, 2, 4], Yuanxing Gui [2], A. Mulero [3]

[1]Shandong Provincial Key Laboratory of Laser Polarization and Information Technology

Department of Physics, Qufu Normal University, Qufu, 273165, P. R. China

[2]Department of Physics, Dalian University of Technology, Dalian, 116024, P. R. China

[3]Department of Applied Physics, University of Extremadura, Badajoz 06072, Spain

[4]Corresponding author, E-mail address: jxtian@dlut.edu.cn

**Abstract:**

Although many equations of state of hard-disk fluids have been proposed, none is capable of reproducing the currently calculated or estimated values of the first eighteen virial coefficients at the same time as giving very good accuracy when compared with computer simulation values for the compressibility factor over the whole fluid range. A new virial-based expression is here proposed which achieves these aims. For that, we use the fact that the currently accepted estimated values for the highest virial coefficients behave linearly with their order, and also that virial coefficients must have a limiting behaviour that permits the closest packing limit in the compressibility factor to be also adequately reproduced.

## 1. Introduction

The study of the hard-disk (HD) fluid has been of great interest for years. As there is no exact theoretical solution for the equation of state (EOS) of this system, a great variety of analytical expressions have been proposed, and new proposals are still being published. The state-of-the-art of this subject up to 2008 has been summarized in a recent book,[1] and a newer short review has been published more recently.[2] In this latter review, it was shown that the most accurate HD EOSs with respect to reproducing the compressibility factor are those proposed by Kolafa and Rottner (KR),[3] which reproduce their own computer simulation results in the density range from 0.4 to 0.89 with an absolute average deviation (AAD) of only 3.5 $10^{-4}$ %. As these EOSs have a large number of fixed and adjustable coefficients, and are thus far from simple. Mulero *et al.* have proposed a clearly simpler expression,[2] called SMC2, which is a modification of the well-known Henderson EOS [4-5]. In particular, the SMC2 EOS gives an AAD=0.10% in the mentioned density range, which is acceptable in view of the expression's simplicity.[2] Unfortunately, the KR and SMC2 proposals cannot reproduce (inside the given uncertainties) the calculated or estimated values obtained by Clisby and McCoy[6] for the first eighteen virial coefficients.

In order to obtain good accuracy at the same time as good virial coefficients, Mulero *et al.* proposed another new EOS,[2] called SMC1, which reproduces the values of the first ten virial coefficients, and gives an AAD=6.4 $10^{-2}$ % in the aforementioned density range. Unfortunately, the predicted values of the virial coefficients higher than the 14th clearly disagree with the estimated values given by Clisby and McCoy,[6] even giving negative values for the 17th and 18th.[7]

All the aforementioned EOSs are based on the use of the virial expansion of the compressibility factor to a certain order, and then to a fit to computer simulation results. Nevertheless, it would be desirable to build new EOSs by using only the virial coefficient values, giving good estimates for the highest ones at the same time as good accuracy for the compressibility factor. Several proposals have been made recently, as will be described in the following.

The behaviour of high-order virial coefficients for the hard-disk fluid has very recently been considered by Hu and Yu.[8] They found that the virial coefficients from third to tenth can be reproduced by using Padé-type approximants. For the higher coefficients, the predictions depend on the Padé-type approximant used, without any clear tendency. They proposed two virial-based

EOSs in which the approximate virial coefficients are used. Comparison with computer simulation data of the compressibility factor was made and good results were obtained. Moreover, in the two new proposals the closest packing density can be explicitly included and then adequately reproduced. Comparison with the estimated values given by Clisby and McCoy[6] for coefficients from the 11th to 18th was not made there.

Also recently, Woodcock and her co-workers[9-10] have also studied the behaviour of virial coefficients for hard disks and the possible construction of new EOSs based on the use of the virial series expansion. In particular, they found that the reduced (relative to the closest packing density) values for the virial coefficients from the twelfth to sixteenth proposed by Kolafa and Rottner[3] can be considered as linear in their order. Although not explicitly mentioned in Ref. 9, this linear relationship was already assumed by Yelash and Kraska[11] several years ago for the first few virial coefficients, and used to propose new simple HD EOSs[1, 11]. In any case, Woodcock[9] proposed a new virial-based EOS for HDs, which seems to give adequate values for the first sixteen virial coefficients and also to be in good agreement with computer simulation data for the compressibility factor. Unfortunately, in this proposal the Clisby and McCoy[6] values for the first ten virial coefficients and their estimates for the eleventh to the eighteenth were not considered. Thus, neither predictions for virial coefficients higher than the 16th nor a comparison with other recently proposed EOSs were made by Woodcock.[9]

Despite the success in reproducing the calculated virial coefficients, the aforementioned EOSs cannot reproduce the values of the coefficients from the eleventh to eighteenth estimated by Clisby and McCoy.[6] Hence, we have recently used[7] the asymptotic expansion method[12] (AEM) to build new EOSs including the first ten virial coefficients. As a result, a new EOS, called AEM(-4, 4) was proposed, which reproduces the first eighteen virial coefficients within the uncertainties given by Clisby and McCoy.[6] Unfortunately, this new EOS gives an AAD of 1.6 % for the fluid density range 0.4–0.89, even though it is only 1.7 $10^{-3}$ % at low densities (from 0.4 to 0.75). In any case, as was noted above, AEM(-4, 4) is the only published EOS that can reproduce the first eighteen virial coefficients within their uncertainties.

In all the aforementioned EOSs, the virial series expansion was taken as the departure point from which to build their analytical expressions. Different approaches were then taken to guarantee the convergence of the series. A consequence is that the only improvement that currently can be made is to construct a new EOS that can reproduce simultaneously the values of the first eighteen virial coefficients and of the compressibility factor. In the present work, we construct such an EOS which is based on the study of the behaviour of the highest known virial coefficients, on the use of one adjustable coefficient to reproduce the compressibility factor data, and on taking into account the limiting behaviour of both the virial coefficients and the compressibility factor at the closest packing density.

The paper is organized as follows: In Sec. 2 we briefly give some details about previous proposals; in Sec. 3 the new proposal is presented, and the new results are compared with those obtained with previous proposals; and finally, Sec. 4 gives the conclusions.

## 2. Virial-Based Equations of State for Hard Disks

As is well-known, the virial series expansion can be considered as the cornerstone of the development of equations of state for fluids. In this expansion, the compressibility factor, $Z$, can be given as an infinite series of terms in powers of the density or packing fraction:

$$Z = \frac{P}{\rho k_B T} = 1 + \sum_{i=2}^{\infty} B_n y^{n-1} \qquad (1)$$

where $P$ is the pressure, $\rho$ the reduced surface density, $T$ the temperature, $k_B$ Boltzmann's constant, and the packing fraction, $y$, is defined in HD fluids as $y = \pi\rho/4$.

The virial coefficients $B_n$ are defined by exact formulas in terms of integrals whose integrands are products of Mayer functions. For hard disks, those integrals are numbers (they do not depend on temperature), but unfortunately only the first four can be calculated analytically and are thus considered as exact values.[1, 13] Values for the virial coefficients from the fifth to the tenth have been obtained by numerical integration by Clisby and McCoy,[6] who also made estimates of the eleventh to the eighteenth. These estimated values are given in the first column of Table 1, where the uncertainty is in the last decimal digit.

Since the virial expansion converges slowly, several types of approximants have been used in order to accelerate the convergence, and then the constructed analytical EOSs are compared with data for $Z$ from computer simulations. The computer simulation data provided by Kolafa and Rottner[3] for reduced densities from 0.4 to 0.89 are currently considered to be the best available, and so are used as referents for the construction or testing of new expressions. It is also important to bear in mind that a continuous-type freezing transition takes place at a density very near[14] to 0.9, so both the computer simulation data and the EOS must have an inflection point at this very high density. At higher densities the geometric properties of the hard-disk molecules demands that the analytical expressions should diverge at the closest packing fraction, $y_c = \pi\sqrt{3}/6 = 0.90689...$, which corresponds to a density of $2/3^{1/2} \approx 1.155$. Thus, the closest packing fraction has been commonly used as parameter with physical significance in a great number of HD EOSs.[1,8-9]

Different approximations to the virial expansion and to the calculated $Z$ values have led to a wide variety of analytical expressions for the EOS of a HD fluid.[1-9,11-12] For instance, three EOSs were constructed by Kolafa and Rottner[3] using the values of the first five virial coefficients and then a fit to both the compressibility factor data, from their own computer simulations, and the higher virial coefficients. Eleven adjustable coefficients were needed in order to give very good accuracy (AAD=3.5 $10^{-4}$% in the density range from 0.4 to 0.89). These three EOSs can therefore be considered as the most accurate at present to give $Z$ values[2], and two of them can reproduce the inflection point at the freezing transition. Nevertheless, as mentioned above, they cannot reproduce the highest known virial coefficients[7].

By studying the behaviour of the virial coefficient values known in 2001, Yelash and Kraska[11] proposed the following approach:

$$B_{n+1} = C_1 + C_2 n + C_0 / y_c^n \qquad (2)$$

which is based on the fact that there is a relationship between the virial coefficient $B_{n+1}$ and the order $n$, and also on the fact that for very high $n$ values the limiting behaviour of virial coefficients must be[15]

$$\lim_{n \to \infty} (B_n / B_{n+1}) = y_c \qquad (3)$$

When Eq. (2) is substituted into the virial expansion, Eq. (1), the sum of the highest terms can be replaced by approximants (Eq. (15) in the Yelash-Kraska paper), resulting in the following expression, which here we call YK:

$$Z_{YK} = 1 + B_2 y + \cdots + B_n y^{n-1} + \frac{C_1 y^n}{1-y} - \frac{C_2 y^n (ny - n - y)}{(1-y)^2} + \frac{C_0 (y/y_c)^n}{1 - y/y_c} \qquad (4)$$

(It has to be noted that the last term of this EOS was misprinted in the original paper of Yelash and Kraska.) Yelash and Kraska[11] used only the values of the first several virial coefficients to derive new simple EOSs, reviewed in Ref. 1. We have reconsidered this new EOS, but using the values for the first eighteen virial coefficients as given by Clisby and McCoy,[6] and then with $n = 15$ and $C_0$, $C_1$, and $C_2$ being obtained from the $B_{16}$, $B_{17}$, $B_{18}$ values using Eq. (2). The values obtained for those coefficients were $C_1 = 8.811616584$, $C_2 = -0.3592865449$, and $C_0 = 2.871576606$. As can be seen in Table 1, the virial coefficients are obviously very well reproduced. Nevertheless, as can be seen in Table 2, the AADs with respect to the computer simulation data[3] for $Z$ are very similar to other proposals, with there being therefore no clear improvement. This means that even reproducing all the presently known values for the virial coefficients, the $Z$ values cannot be simultaneously reproduced with accuracy if no adjustable parameters are introduced. This is due to the uncertainties and statistical errors in both computer simulation data and virial coefficients calculations.

More recent proposals take into account the behaviour of the virial coefficients known at present. In particular, Hu and Yu[8] have proposed the following approximation to calculate the virial coefficients higher than the second:

$$B_n = n + \Delta B_n, n \geq 3 \tag{5}$$

where $n$ is the order of the coefficient, and the values $\Delta B_n$ are represented by a Padé-type approximant. The most accurate approximant constructed was a [4/3] one, including the values of the first nine virial coefficients. The $B_n$ values obtained for $n>10$ are compared in Table 1 with those estimated by Clisby and McCoy.[6] As can be seen, the approximation is valid only for $n<13$.

Two new HD EOSs were proposed by Hu and Yu, based on Eq. (3). The first proposal, which we shall call HY1, is to sum the infinite sequence from order higher than a given $m$ value, as follows:

$$Z_{HY1} = \sum_{n=1}^{m} B_n y^{n-1} + \frac{B_{m+1} y^m}{1 - y/y_c} \tag{6}$$

The second proposal is to include the limiting behaviour together with the second virial coefficient:

$$Z_{HY2} = 1 + \frac{B_2 y}{1 - y/y_c} + \sum_{n=3}^{m} \left( B_n - B_2 / y_c^{n-2} \right) y^{n-1} \tag{7}$$

Both expressions include the closest packing density and they predict the same values for the highest virial coefficients, as given in Table 1, as they use the same Padé-type approximant for $\Delta B_n$ in Eq. (5). As can be seen in Table 2, these new EOSs (with $m = 18$) give similar AADs, in different density ranges, as given by other EOSs, some of which give better values of the highest known virial coefficients.

Recently Woodcock and coworkers[9,10] have considered the virial expansion by including the $y_c$ value explicitly, and then studied the behaviour of $B_n y_c^{n-1}$ versus $n$. By using the Kolafa and Rottner[3] data, they found the relationship:

$$B_n y_c^{n-1} = C - An, \text{ for } n > 11 \tag{8}$$

where $C$ and $A$ are adjustable parameters that were obtained by Woodcock using the values of $B_{12}$ to $B_{16}$. This relationship is a special case of that previously proposed by Yelash and Kraska[11], Eq. (2), and it permits one to write the higher terms in the virial expansion as a closed form, as follows[9]:

$$Z_W = 1 + \sum_{n=2}^{m} B_n y^{n-1} + \left(\frac{y}{y_c}\right)^m \left[\frac{C - mA}{1 - y/y_c} - \frac{A}{(1 - y/y_c)^2}\right] \quad (9)$$

We have again reconsidered here the relationship (8) and Eq. (9), but using the Clisby and McCoy[6] values for the virial coefficients. As can be seen in Fig. 1, the linearity is maintained for $n>12$ with this new data, so we found the new $C$ and $A$ values by using the values from $B_{13}$ to $B_{18}$ estimated by Clisby and McCoy[6], obtaining: $A$=0.130 213 158 443 84 and $C$=5.734 384 107 471 29.

As can be seen in Table 1, very good values are obtained for the highest known virial coefficients, clearly better than those given by Hu and Yu.[8] Nevertheless, when the new Woodcock EOS, Eq. (9), is used with $m = 12$, the AAD values obtained with respect to the $Z$ data are very similar to those given by the HY1, HY2, and other similar EOSs.

In sum, the above results show that the aforementioned methods need to include at least the $B_{12}$ value explicitly in the EOS to obtain adequate predictions for the 13th to 18th virial coefficients. Nevertheless, we have very recently shown[7] that those high virial coefficients can be reproduced by using only the values of the first ten. For that, we used the asymptotic expansion method (AEM)[12] to build a new EOS that does not include any adjustable coefficient. The AEM EOS can be written as:

$$Z_{AEM}(i, j) = \sum_{k=i}^{j} a_k (y - b_{ij})^{-k} \quad j > i; i, j, k \in N \quad (10)$$

where $a_k$ are coefficients to be determined, $b_{ij}$ is the radius of convergence of the virial expansion which must be greater than $y_c$, and $k$ can take both positive and negative integer values. Both the $a_k$ and $b_{ij}$ values are determined by comparison with the virial expansion, Eq. (1). After considering 57 possible EOSs in the form of Eq. (10), we showed[7] that only the choice $i=-4$, $j=4$ with $b_{ij}=$ 1.061330772 can give the first eighteen virial coefficient values within the uncertainties given by Clisby and McCoy[6]. The $a_k$ values can be obtained by using the values of the first ten virial coefficients and can be found in Ref. 7. The predicted virial coefficients obtained using the AEM(-4,4) EOS are given in Table 1. Nevertheless, as one observes in Table 2, even this new expression cannot give a low AAD at high densities for the $Z$ values. Moreover, the $b_{ij}$ value obtained is significantly higher than any other value with physical significance (the freezing density, the $y=1$ reference used in a great number of EOSs, or the closest packing density). We have also tried to obtain a fitted value for $b_{ij}$ in order to obtain a clearly lower AAD by using Eq. (10), but unfortunately in this case the highest virial coefficients could not be reproduced. So, it seems that no improvement can be got with the AEM until new virial coefficients have been calculated in the future.

In the present work, our proposal is to reconsider the Yelash and Kraska[11] relationship, Eq. (2), for representing the behaviour of the highest virial coefficients. We also note that, although not explicitly written, the Woodcock proposal, Eq. (9), is a particular case of the YK expression. Also, the HY expressions, Eqs. (6-7), are based on Eq. (2) given by Yelash and Kraska[11].

## 3. A New Virial-Based EOS

Our new virial-based EOS is based on the virial expansion, and on the Yelash-Kraska EOS, Eq. (4). We consider separately the first $n-1$ terms of the virial expansion, then the terms from $n$ to $m$, which are taken to be linear with their order, and then the remaining terms, which are taken to be appropriate to give the correct pole in the EOS. Thus, the HD EOS is written as:

$$Z = Z_T + Z_L + Z_I \tag{11}$$

where

$$Z_T = 1 + B_2 y + B_3 y^2 + \cdots + B_n y^{n-1} \tag{12}$$

$$Z_L = \sum_{i=n}^{m}(c_1+c_2 i)y^i \tag{13}$$

$$Z_I = \sum_{j=m+1}^{\infty} c_0 y^j / y_c^j \tag{14}$$

In the above expressions, $Z_T$ is the truncated virial series to $(n-1)$-th order. $Z_L$ is used to take account of the virial terms of which the virial coefficients $B_{n+1}$ to $B_{m+1}$ are linear with their order, and $Z_I$ is designed to let the limiting behaviour of the virial coefficients of Eq. (11) satisfy Eq. (3).[11]

The closed form of Eq. (11) is solvable as follows:

$$\begin{aligned}Z = 1 + B_2 y + B_3 y^2 + \cdots + B_n y^{n-1} + \\ \frac{(c_1+c_2 n)y^n + (c_2-c_1-c_2 n)y^{n+1} + (-c_2-c_1-c_2 m)y^{m+1} + (c_1+c_2 m)y^{m+2}}{(1-y)^2} \\ + \frac{c_0 y^{m+1}}{y_c^{m+1}(1-y/y_c)}\end{aligned} \tag{15}$$

In our calculations, we choose $n=10$ in order to use the calculated[6] first ten virial coefficients. The truncated $Z_T(n=10)$ EOS gives good $Z$ values only for $y$ below 0.55, as can be seen in Fig. 3. Therefore, a contribution from higher virial coefficients is needed, which can be done by using the $Z_L$ term.

Figure 2 plots the mean $B_{11}$ to $B_{18}$ values, as given by Clisby and McCoy[6], versus their order $n$. As can be seen, the linear relationship included in Eq. (13) stands up well for $B_{11}$-$B_{17}$. That $B_{18}$ is not located on the straight line may be due to its large relative error of the estimate given by Clisby and McCoy.[6] Obviously, if the statistical errors in the virial coefficients were also plotted in Fig. 2, we could see that some other relationship, and not only the linear one, could be considered. Nevertheless, the proposed linear relationship is the simpler one.

Considering only $B_{11}$ to $B_{17}$, $c_1$ and $c_2$ take the values $c_1$=1.74423771428571; $c_2$=0.93981257142857. Using these values, the $Z_T$+$Z_L$($n=10$, $m = 18$) EOS can be constructed with the guarantee that it reproduces the first eighteen virial coefficients, as is seen in Table 1. As can be seen in Table 2, the AADs obtained are similar to those of other EOSs constructed using the presently known values for the first eighteen virial coefficients.

We then considered higher values for $m$, and found that the minimum AAD1 value with respect to the Kolafa and Rottner $Z$ values[3] at densities from 0.4 to 0.75 occurs for $m = 22$. This means that in our proposal the virial coefficients from $B_{11}$ to $B_{22}$ are taken to be linear with their order. At higher values of $m$, the AADs increase in the three density ranges defined in Table 2. When values of $n$ higher than 10 are considered, the minimum AAD1 value is still obtained for $m = 22$ without any significant improvement over the choice $n = 10$.

As can be seen in Tables 1 and 2, when only the terms $Z_T+Z_L$ ($n=10$, $m=22$) are considered, the first eighteen virial coefficients are reproduced, but again the AADs obtained are similar to other previous EOSs. In particular, this new EOS gives adequate Z values only for $y \leq 0.65$ (Fig. 3), overestimating the values at higher densities.

As a conclusion, the results obtained with this new $Z_T+Z_L$ EOS, and also with the previous AEM(-4,4) and YK EOSs, confirm that the use of virial coefficients alone cannot give low AADs, although the first eighteen virial coefficients are reproduced within their uncertainties. This is due to the fact that the statistical errors obtained in the calculation of both virial coefficients and Z values from computer simulations, are added to the simplifications included in the EOSs.

Thus, the only way to reduce the AADs at the same time as giving the correct virial coefficients and density pole seems to be to include at least one adjustable coefficient. We thus considered the three terms in Eq. (15) with $c_0$ being obtained from a fit to the Kolafa and Rottner[3] data in the density range from 0.80 to 0.89 (i.e., by minimizing the AAD2 value in Table 2). We made calculations with $m$ values from 22 to 100, and found that the lowest AAD2 value is obtained for *n=10, m=45*, and *c₀=-40233.04* in Eq. (15), which leads to an AAD3 for the entire density range of only 3.3 $10^{-2}$% (see the agreement with data in Fig. 3). This therefore implies a clear improvement with respect to the other EOSs reproducing the first eighteen virial coefficients known at present. Moreover, as can be seen in Fig. 3, this fit permits also reproduce the inflection point at very high densities, as the KR EOSs do.

## 4. Conclusions

Several recent proposals for the equation of state of hard-disk fluids have been considered and analyzed in order to reproduce the calculated first ten virial coefficients and the estimated values published by Clisby and McCoy[6] for the 11th to the 18th, at the same time as having good accuracy in the calculation of the compressibility factor at densities below 0.89.

All the equations of state analyzed or proposed were based on the virial expansion, and were considered together and then compared for the first time in this study. In particular, we have shown that the asymptotic expansion method is the only method of those analyzed here that can give an equation of state reproducing the estimated values of the first eighteen virial coefficients from knowledge of only the values of the first ten. All the other methods considered here to accelerate the convergence of the series, and to yield adequate values of the virial coefficients, require knowledge of at least the values of the first twelve.

We have shown that when an EOS is constructed using only virial coefficient values up to 18th order, with no adjustable parameters, it cannot adequately reproduce the compressibility factor at high densities, even if all the known values of the first eighteen virial coefficients are used in constructing the EOS. This is due to the uncertainties in computer simulations, virial coefficients calculations, and to the used simplifications in EOSs.

A new analytical expression for the equation of state of hard-disk fluids has been proposed. It comprises three terms: (i) a truncated virial series including the values of the first ten virial coefficients; (ii) a second term in which the virial coefficients from the 11th to the 45th are taken to be linear with their order (the two parameters needed for the linear relationship were determined using the presently known estimated values of the 11th to the 17th virial coefficients); and (iii) an asymptotic term that takes the limiting behaviour of virial coefficients into account. This last term requires a fit to the values of the compressibility factor given by computer simulations – in the present proposal this is done with only one adjustable parameter.

We then showed that this new equation of state is the only one presently published that can reproduce the known values of the first eighteen virial coefficients at the same time as giving only small deviations for the compressibility factor values over the entire density range. The main limitation of the present EOS is the same as all the previous ones: it cannot be extrapolated with guarantees at densities higher than 0.9 because of the freezing fluid-to-hexatic transition.

Finally we would stress that the proposals considered here are based on what are presently considered the best estimates of the 11th to the 18th virial coefficients. While these coefficients may be calculated more accurately in the near future, the resulting values are unlikely to be very different from those currently known, so that their linearity with respect to their order will be maintained. As long as this is the case, the new values can be incorporated into the proposed EOS with no difficulty. As has been shown in the present study, other methods may require major modifications and/or the inclusion of further adjustable coefficients.

## Acknowledgements

The National Natural Science Foundation of China under Grant No. 10804061, the Natural Science Foundation of Shandong Province under Grant No. Y2006A06, and the foundations of QFNU and DUT have supported this work (J.T. and Y.G.).

**Table Captions**

**Table 1** Estimated values of the eleventh to eighteenth virial coefficients for hard-disk fluids. YK = Eq. (4) with $n$ =15. HY = Eqs. (5-7) with $\Delta B_n$ being obtained by a [4/3] Padé-type approximant given in Ref. 8. W = Eqs. (8-9) with parameters calculated here and given in the text. AEM(-4,4) = Eq. (10) with parameters given in Ref. 7. This work = Eq. (15) with parameters given in the text obtained by using the $B_{11}$ to $B_{17}$ values. Cursive and bold values mean that they have been explicitly included in the equation of state.

**Table 2** Location of the first pole and absolute average deviations (AADs) of the compressibility factor from EOSs when compared with the computer simulation results of Kolafa and Rottner[3] over different density ranges: AAD1 ($\rho$=0.4-0.75), AAD2 ($\rho$=0.8-0.89), AAD3 ($\rho$=0.4-0.89). YK, HY1, HY2, and W, are Eqs. (4), (6), (7), and (9), respectively, with the parameter values given in the text, which may be different from those originally proposed in the references included. AEM(-4,4) = Eq. (10) with parameters given in Ref. 7. $Z_T+Z_L$ = Eq. (15) with $n$=10, and $m$ = 18 or $m$=22, and $c_0$=0. ($Z_T+Z_L+Z_I$) = Eq. (15) with $n$=10, $m$=45 and $c_0$=-40233.04.

**Figure Captions**

**Figure 1** Plot of virial coefficients multiplied by the closest packing fraction, $B_n y_c^{n-1}$, versus their orders. Line: linear fit; points: data from Clisby and McCoy.[6]

**Figure 2** Plot of $B_{n+1}$ versus their order. Points: $B_{11}$~$B_{18}$ by Clisby and McCoy[6]; line: the linear fit using only $B_{11}$~$B_{17}$ values.

**Figure 3** Plot of the compressibility factor $Z$ versus the packing fraction $y$ from the computer simulation of Kolafa and Rottner[3] and several EOSs based on Eq. (15).

Table 1 Estimated values of the eleventh to eighteenth virial coefficients for hard-disk fluids. YK = Eq. (4) with $n$ =15. HY = Eqs. (5-7) with $\Delta B_n$ being obtained by a [4/3] Padé-type approximant given in Ref. 8. W = Eqs. (8-9) with parameters calculated here and given in the text. AEM(-4,4) = Eq. (10) with parameters given in Ref. 7. This work = Eq. (15) with parameters given in the text obtained by using the $B_{11}$ to $B_{17}$ values. Cursive and bold values mean that they have been explicitly included in the equation of state.

| Virials | Ref. 6 | YK | HY | W | AEM(-4,4) | This work |
|---|---|---|---|---|---|---|
| $B_{11}/B_2^{10}$ ($10^{-2}$) | 1.089 | ***1.089*** | 1.0899… | *1.089* | 1.0894… | 1.0881… |
| $B_{12}/B_2^{11}$ ($10^{-3}$) | 5.90 | ***5.90*** | 5.908… | *5.90* | 5.904… | 5.899… |
| $B_{13}/B_2^{12}$ ($10^{-3}$) | 3.18 | ***3.18*** | 3.412… | 3.187… | 3.179… | 3.179… |
| $B_{14}/B_2^{13}$ ($10^{-3}$) | 1.70 | ***1.70*** | 1.820… | 1.700… | 1.703… | 1.704... |
| $B_{15}/B_2^{14}$ ($10^{-4}$) | 9.10 | ***9.10*** | 9.676… | 9.065… | 9.083… | 9.095… |
| $B_{16}/B_2^{15}$ ($10^{-4}$) | 4.84 | 4.839… | 5.124… | 4.825… | 4.823… | 4.834… |
| $B_{17}/B_2^{16}$ ($10^{-4}$) | 2.56 | 2.559… | 2.705… | 2.565… | 2.551… | 2.560… |
| $B_{18}/B_2^{17}$ ($10^{-4}$) | 1.36 | 1.359… | 1.089… | 1.362… | 1.344… | 1.352… |

**Table 2** Location of the first pole ($y_{pole}$ giving Z = 0) and absolute average deviations (AADs) of the compressibility factor from EOSs when compared with the computer simulation results of Kolafa and Rottner[3] over different density ranges: AAD1 ($\rho$=0.4-0.75), AAD2 ($\rho$=0.8-0.89), AAD3 ($\rho$=0.4-0.89). YK, HY1, HY2, and W, are Eqs. (4), (6), (7), and (9), respectively, with the parameter values given in the text, which may be different from those originally proposed in the references included. AEM(-4,4) = Eq. (10) with parameters given in Ref. 7. $Z_T+Z_L$ = Eq. (15) with $n=10$, and $m = 18$ or $m=22$, and $c_0=0$. ($Z_T+Z_L+Z_I$) = Eq. (15) with $n=10$, $m=45$ and $c_0=-40233.04$.

| EOS | $y_{pole}$ | AAD1 (%) | AAD2 (%) | AAD3 (%) |
|---|---|---|---|---|
| KR [3] | 1 | 1.3 E-4 | 5.6 E-4 | 3.5 E-4 |
| SMC2 [2] | 1 | 0.12 | 8.6 E-2 | 0.10 |
| SMC1 [2] | 0.8391 | 3.9 E-2 | 8.8 E-2 | 6.4 E-2 |
| YK [11], n=15 | 0.9069 | 2.2 E-3 | 3.2 | 1.6 |
| HY1 [8] | 0.9069 | 2.3 E-3 | 3.2 | 1.6 |
| HY2 [8] | 0.9069 | 3.3 E-3 | 3.0 | 1.5 |
| W [9] | 0.9069 | 2.0 E-3 | 3.2 | 1.6 |
| AEM(-4,4) [7] | 1.0613 | 1.7 E-3 | 3.1 | 1.6 |
| $Z_T+Z_L$ ($m$=18) | 1 | 6.5 E-3 | 2.7 | 1.4 |
| $Z_T+Z_L$ ($m$=22) | 1 | 1.4 E-3 | 3.0 | 1.5 |
| $Z_T+Z_L+Z_I$ | 0.9069 | 1.3 E-3 | 6.5 E-2 | 3.3 E-2 |

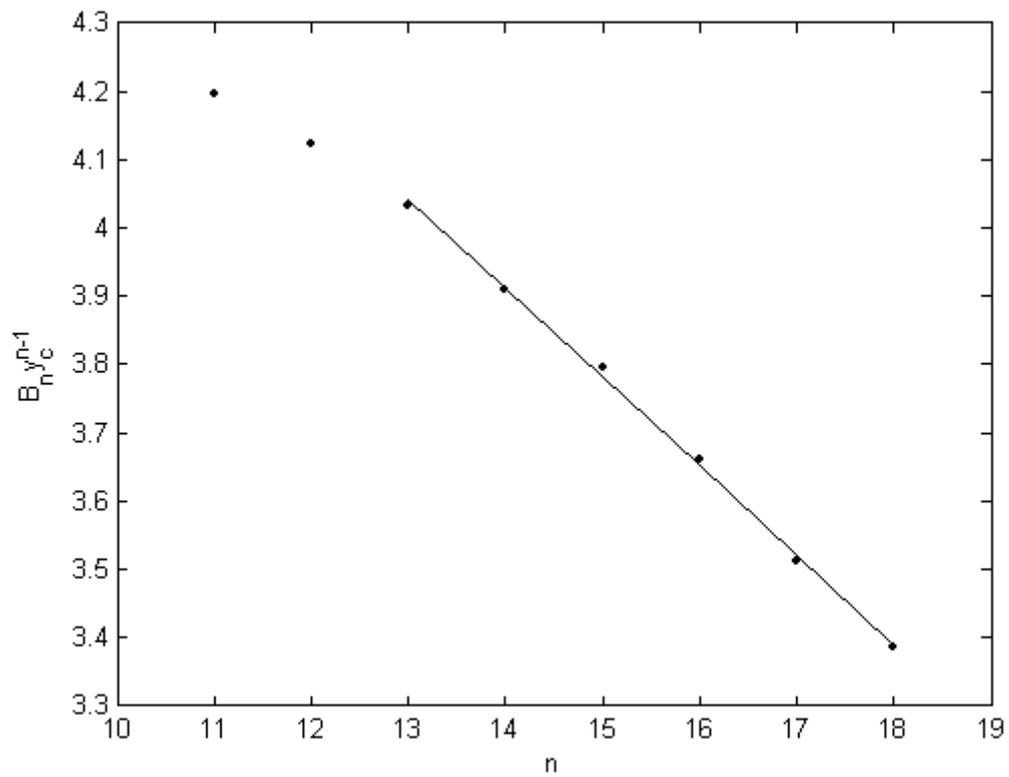

**Figure 1** Plot of virial coefficients multiplied by the closest packing fraction, $B_n y_c^{n-1}$, versus their orders. Line: linear fit; points: data from Clisby and McCoy.[6]

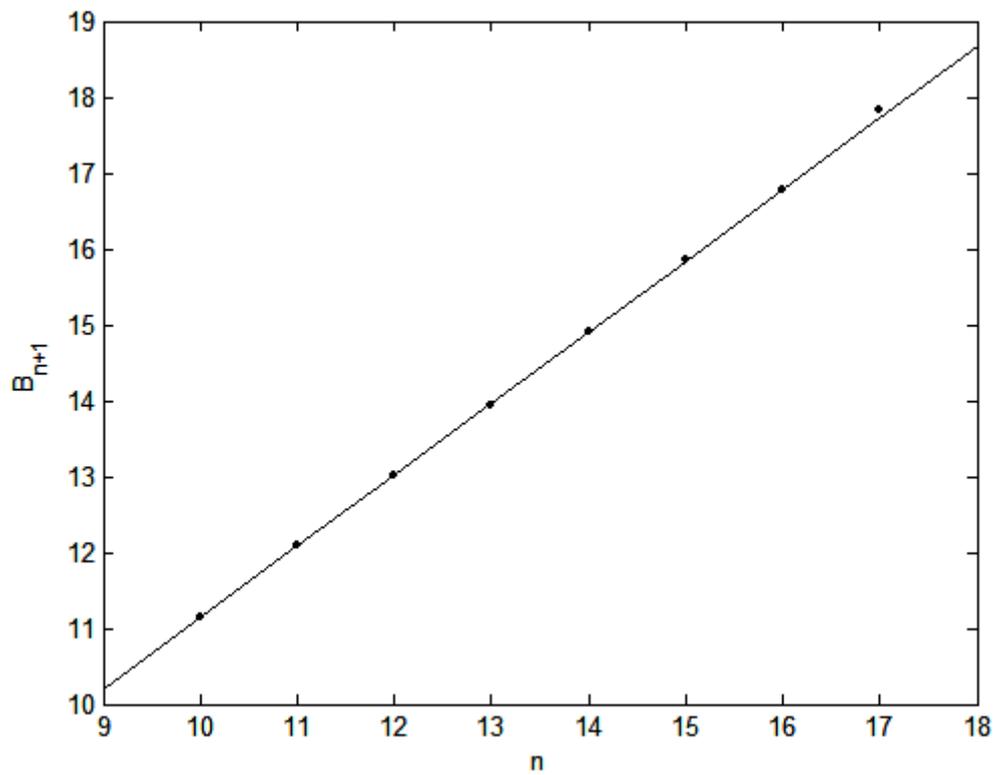

**Figure 2** Plot of $B_{n+1}$ versus their order. Points: $B_{11}$~$B_{18}$ by Clisby and McCoy[6]; line: the linear fit using only $B_{11}$~$B_{17}$ values.

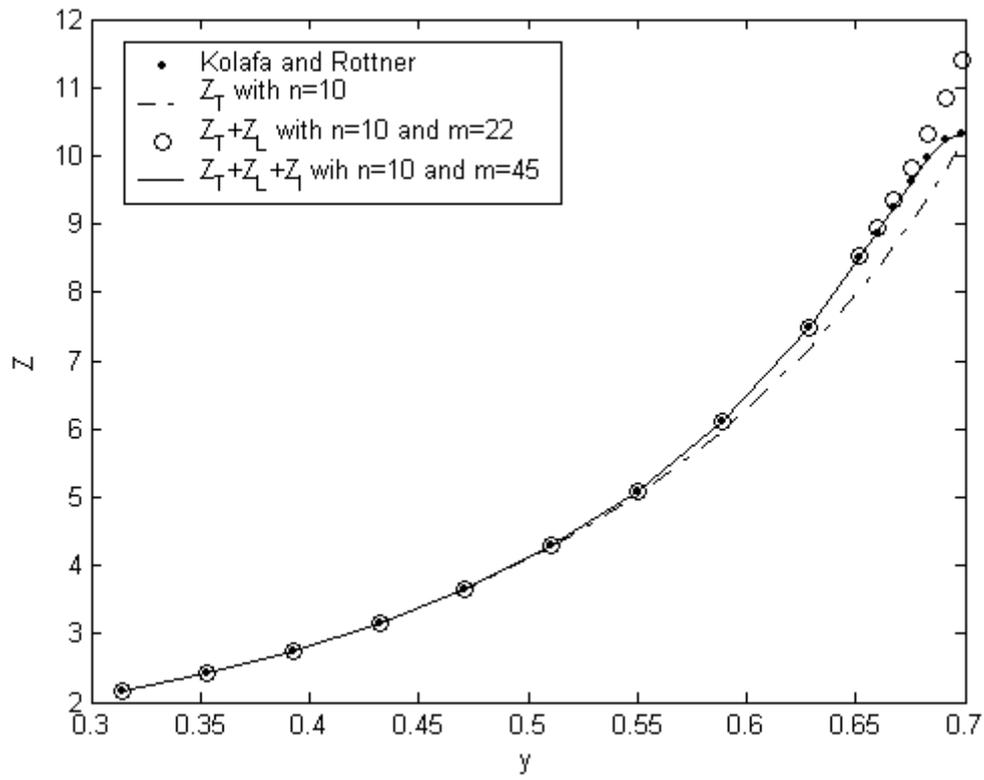

**Figure 3** Plot of the compressibility factor $Z$ versus the packing fraction $y$ from the computer simulation of Kolafa and Rottner[3] and several EOSs based on Eq. (15).